\documentstyle[12pt]{article}

\begin{document}
\baselineskip=18.6pt plus 0.2pt minus 0.1pt \makeatletter
\@addtoreset{equation}{section} \renewcommand{\theequation}{\thesection.%
\arabic{equation}} \makeatletter \@addtoreset{equation}{section}
\date{}
\begin{titlepage}

\title{\vspace{-3cm}
\hfill\parbox{4cm}{}\\
 \vspace{1cm}
 Implications of dilaton couplings on the axion potential}
 \vspace{2cm}

\author{N. El Biaze$^{1,2}$\footnote{n.elbiaze@uiz.ac.ma}, R. Markazi$^{1,2}$\footnote{r.markazi@uiz.ac.ma} \\
\
\small{$^1$ Materials and renewable energies laboratory, Faculty of science, Ibn Zohr University-Agadir, Morocco.}
\\ \small{$^2$ High School of Technology-Guelmim, Ibn Zohr University-Agadir, Morocco.}} \maketitle \setcounter{page}{1}
\begin{abstract}
 In particle physics and cosmology, the dilaton is an hypothetical scalar field which can explain many physical phenomena. In this framework, we investigate an extended lagrangian of QCD which involves dilatonic degrees of freedom. Our approach is based on the relationship between the massive dilaton and the nonperturbative effects of QCD. We derive a general axion potential involving the dilaton properties and vacuum condensates.

\end{abstract}
\end{titlepage}

\newpage
\section{Introduction}
\label{intro}
{The axion was introduced by Weinberg
{[}1{]} and Wilczek {[}2{]} to solve the strong CP problem in QCD
at very short distances. This is} due to the presence of the topological
violating P and CP term $\theta\frac{g^{2}}{32\pi}G_{{\mu}\nu}\tilde{G_{{\mu}\nu}}${[}4{]},
where $G_{{\mu}\nu}$ is the QCD field strength, $\tilde{G_{{\mu}\nu}}$
its dual and $\theta$ denotes the usual QCD vacuum angle.{{}
Through the Peccei-Quinn mechanism, this particle was also introduced
as a pseudo-Goldstone boson resulting from the spontaneous breakdown
of the $U(1)_{PQ}$ global symmetry.}Furthermore, this particle is a candidate for dark matter in cosmology
and may constitute the missing mass of universe {[}3{]}. It can be
also defined as a scalar field which appears in several theories.
In supersymmetric gauge theories, the axion is defined as a companion
for a particle called dilaton. This latter plays crucial roles in
particles physics and cosmology where it can be either massive or
masseless. The real part of the lowest component of a chiral superfield
corresponds to the dilaton field, whereas the imaginary part is the
axion field.

{Within the Kaluza-Klein supergravity frame, the dilaton
arises from the $d_{i}$ internal dimensions. In fact, if the field content
of space-time is assumed to arise from embedding in $(4+d_{i})$-dimensional
manifold, the $d_{i}$ internal dimensions will have variable volume
from the four-dimensional point of view.} In string theory, the axion
and the dilaton appear in the masseless spectrum of $10D$ type $IIB$
model and its lower dimensional compactifications as well{[}4{]}.
From the previous descriptions of the dilaton, we see that there is
no unique definition of this particle because it interacts in various
ways. For example, it couples naturally to super-Yang Mills gauge
fields in curved space and plays a central role in string theory since
it defines the string coupling constant $g_{s}$ as $e^{\phi}$. In
Brans-Dicke model of induced gravity, it appears in the coupling
dilaton-Ricci curvature term whereas in the QCD lagrangian, it appears
in the coupling between the dilaton and the field strength. This coupling
can generate an interquartk potential{[}8{]},
which exhibits both a short range Coulomb potential and a linear confining
term at large distances.
In order to derive a general form for the axion potential, we shall
extend the Vecchia, Veneziano and Witten (VVW) {[}5, 6{]} model,
of standard QCD in the chiral limit and for large colour number $N_{C}$
by taking into account new possible dilaton interactions.
\section{Review on the axion potential}
{We recall that, near the chiral limit $m_{q}\rightarrow0$
with equal masses, the exact Ward Identity (WI) takes the form {[}6{]}:}

\begin{equation}
{i\int dx<0||T{\frac{\alpha}{8\pi}G\tilde{G}(x),\frac{\alpha}{8\pi}G\tilde{G}(0)}||0>=-\frac{\partial^{2}Evac(\theta)}{\partial\theta^{2}}}
\end{equation}

{From this equation, one can easily show that the vacuum
energy $E_{vac}(\theta)$ dependence on the $\theta$-angle appearing
in the CP-violating term of QCD lagrangian:}

\begin{equation}
E_{vac}(\theta)=E(0)-\frac{1}{2}\chi\theta^{2}
\end{equation}

where $\chi$ is the topological susceptibility given by the following
formula{[}7{]}:
\begin{equation}
\chi=i\int dx<0||T{\frac{\alpha}{8\pi}G\tilde{G}(x),\frac{\alpha}{8\pi}G\tilde{G}(0)}||0>=\frac{m_{q}}{N_{f}}<0|\bar{\psi}\psi||\bar{\psi}\psi|0>+O(m_{q}^{2}), \end{equation}

with $N_{f}$ is the number of flavors and $<0|\bar{\psi}\psi|0>$represents
the quark condensate.

{We know that the vacuum energy $E_{vac}(\theta)$
is a periodic function of $\theta$, i.e. $E_{vac}(\theta+2\pi)=E_{vac}(\theta).$
This periodicity is a direct consequence of the quantization of the
topological charge in QCD: $\left|\theta_{i}\right\rangle =\sum e^{in\theta}\left|n\right\rangle $,
$\left|\theta+2\pi\right\rangle \equiv\left|\theta\right\rangle $.}
By using the one-to-one correspondence $\mathbf{\mathrm{\theta\rightarrow}\mathrm{\frac{a}{f_{a}}}}$
$[11${]}, we can derive the following axion potential:

\begin{equation}
V(a)=V(0)-\frac{1}{2}\chi(\frac{a}{f_{a}})^{2}
\end{equation}

where $f_{a}$ denotes the axion coupling constant.

The equation (2.4) is unable to describe the periodic behaviour of
$\theta$-angle, which is an essential property for the investigation
of vacuum features mainly the metastable states and the domain walls
{[}8{]}.
{In gluodynamics with large number of colors
$N_{c}$, the topological term of the lagrangian should depend on
$\theta/N_{c}$ in order to solve the $U(1)$ problem {[}9{]}.}
One can exhibit this periodicity in $\theta$ by considering the Di
Vecchia-Venezino-Witten (VVW) model of standard QCD. Their lagrangian
in the chiral limit and for large colour number $N_{c}$ {[}5, 6{]}
becomes:

\begin{equation}
L=\frac{F\pi}{2}Tr(\partial_{\mu}U\partial^{\mu}U+)+Tr(MU+M^{+}U^{+})+\frac{c}{N_{c}}(-ilog(detU)-\theta)^{2}+....
\end{equation}

where $M=diag(m_{i}<$$\bar{\psi^{i}}\psi^{i}>)$, $\frac{c}{N_{c}}$
is the topological suscepibility and $U$ matrix is the
diagonal parameterization of the linear $\sigma$-model,

\begin{equation}
U=\left(\begin{array}{ccc}
e^{-i\varphi_{1}} & 0 & 0\\
0 & e^{-i\varphi_{2}} & 0\\
0 & 0 & e^{-i\varphi_{3}}
\end{array}\right)
\end{equation}
The WI anomalous  requires {[}10{]} that Goldstone
bosons which are described by the unitary matrix $U$ should appear
only in the following combination with $\theta$: $\theta-iTrlogU$.

{In order to study the vaccum energy $E_{vac}(\theta)$
in the chiral limit and for large value of colour number $N_{c}$,
Di Vecchia, Veneziano and Witten {[}10{]} suggested the following
effective potential:}
\begin{equation}
W_{VVW}(\theta,U)=-\frac{\chi}{2}(\theta-ilogDetU)^{2}-\frac{1}{2}Tr(MU+M^{+}U^{+})+....
\end{equation}

{By including the missing terms at order up to next-to-leading order 
and making the correspondance: $\theta\rightarrow\frac{a}{f_{a}}$,
one can derive the following potential:}
\begin{equation}
V(a)=-b<0|\frac{\alpha_{s}}{32\pi}G^{2}||0> \\ +m_{q}N_{c}<0|\bar{\psi}\psi||0>cos(\frac{a}{f_{a}N_{c}})
\end{equation}

\section{Axion potential with dilatonic degrees of freedom}
We extend the effective lagrangian of VVW model {[}5, 6{]} by introducing
the dilaton degrees of freedom as follows:
$$L_{VVW}=L_{0}(U,U^{+})+L(A_{\mu},\phi,\theta)+\frac{1}{8}i\varepsilon_{\mu\nu\rho\sigma}G^{\mu\nu\rho\sigma}Tr[logU-logU^{+}] $$
\begin{equation}
+\frac{f_{\pi}}{2{{ \sqrt{2}}}}Tr[MU+MU^{+}]
\end{equation}
At large $N_{c}$ limit, in which QCD exhibits both confinement and
chiral symmetry breaking, the lagrangian (3.1) may be written as a
sum of the following terms:
$L_{0}(U,U^{+})$ is a sigma model lagrangian which describes the
low energy dynamics of the pseudoscalar mesons parameterized by the
$U$ complex matrix:
\begin{equation}
L_{0}(U,U^{+})=\frac{1}{2}Tr(\partial_{\mu}U\partial^{\mu}U^{+}).
\end{equation}
$L(A_{\mu},\phi,\theta)$ which depends
on the gauge field $A_{\mu}$, the dilaton $\phi$
and the $\theta-$vacuum angle as follows:
\begin{equation}
L(A_{\mu},\phi,\theta)=\frac{d}{F(\phi)}G_{\mu\nu\rho\sigma}G^{\mu\nu\rho\sigma}-\frac{\theta}{4}\epsilon_{\mu\nu\rho\sigma}G^{\mu\nu\rho\sigma} \\
-\frac{1}{2}\partial_{\mu}\phi\partial^{\mu}\phi-\frac{1}{2}m^{2}\phi^{2},
\end{equation}
Where the $\theta$-term in the QCD fundamental lagrangian was substituted
by: $-\frac{\theta}{4}\varepsilon_{\mu\nu\rho\sigma}G^{\mu\nu\rho\sigma}$.
The tensor $G_{\mu\nu\rho\sigma}$ is the field strength
for a three-index field $A_{\mu\nu\rho}$ which depends
on the gauge field $A_{\mu}$ as follows:
$$A_{\nu\rho\sigma}=\frac{g^{2}}{96\pi^{2}}[A_{a \nu}{{ \overleftarrow{{\partial}}}}_{\rho}A_{\sigma}^{a}+A_{a \nu}{ {\partial}}_{\rho}A_{\sigma}^{a} {+}A_{a \rho} { { \overleftarrow{{\partial}}}}_{\nu}A_{\sigma}^{a}-A_{a \rho}\partial_{\nu}A_{\sigma}^{a}$$

\begin{equation}
-A_{a \nu} \overleftarrow{{\partial}}_{\sigma}A_{\rho}^{a}-A_{a \nu}\partial_{\sigma}A_{\rho}^{a}+2f_{abc}A_{\nu}^{a}A_{\rho}^{b}A_{\sigma}^{c}]
\end{equation}
$f_{abc}$ are the $SU(N_{c})$ structure constants.

Since the massive dilaton couples to the YM term like in Dick's model,
we may propose that the kinetic term $G_{\mu\nu\rho\sigma}G^{\mu\nu\rho\sigma}$
of the field $A_{\nu\rho\sigma}$ in VVW model couples to the dilaton
field $\phi$ via this factor $\frac{d}{F(\phi)}$.
Where $d$ is a positive constant and $m$ is the dilaton mass.

If we consider the field $q(x)=\frac{1}{4}\varepsilon_{\mu\nu\rho\sigma}G^{\mu\nu\rho\sigma}$,
we can rewrite the effective lagrangian at large-$N_{c}$ as:
$$L(U,U^{+},q,\phi)=L_{0}(U,U^{+})+\frac{1}{F(\phi)}N_{c}q^{2}(x)-\theta q(x)+\frac{1}{2}iq(x)Tr[logU-logU^{+}]$$
\begin{equation}
+ \frac{f_{\pi}}{2 \sqrt{2}}Tr[MU+MU^{+}]-\frac{1}{2}\partial_{\mu}\phi\partial^{\mu}\phi-\frac{1}{2}m^{2}\phi^{2}
\end{equation}
with $c=\frac{4!}{d}$.
By using the equation of motion of $q(x)$, we derive the following
effective potential $W$ which depends both on $\theta$ and $\phi$:
\begin{equation}
W(\theta,\phi)=W(\theta=0,\phi)-\frac{cF(\phi)}{4N_{C}}(\theta {+}\frac{1}{2}iTr[logU-logU^{+}])^{2}-\frac{1}{2}m^{2}\phi^{2}.
\end{equation}
We notice that the two first terms of equation (3.6) represent
the cosine expansion.
If we take in consideration the periodic behaviours of the $\theta$-angle,
we can get a more general form for the potential $W(\theta,\phi)$,
namely:
\begin{equation}
W(\theta,\phi)=\frac{cF(\phi)}{4}N_{C}cos(\frac{1}{N_{C}}(\theta-\sum\varphi_{i}))-\frac{1}{2}m^{2}\phi^{2}.
\end{equation}
Therefore, via $\theta\rightarrow\frac{a}{f_{a}}$ correspondence
and the dilatonic vacuum expectation value, the axion potential can
be written like:
\begin{equation}
W(\theta,\phi)=\frac{c<F(\phi)>}{4}N_{c}cos(\frac{1}{N_{C}}(\frac{a}{f_{a}}-\sum\varphi_{i}))-\frac{1}{2}m^{2}\phi^{2}.
\end{equation}
this form (3.8) is more general because it allows us to describe
the axion potential for any kind of dilaton-gluon coupling $F(\phi)$.
In the particular case: $F(\phi)=\frac{\phi}{f}$, the dilaton mass
and the vacuum expectation value $<\phi^{2}>$ can be substituted by using the following relations
$[12]$:
\begin{equation}
m^{2}f_{\phi}^{2}=-\frac{2\beta(g)}{g}<0|G_{\mu\nu}G^{\mu\nu}|0>
\end{equation}
$$4<\phi^{2}>=-\frac{g}{\beta(g)}\frac{f_{\phi}^{3}}{f}-f_{\phi}^{2}$$

with $\frac{-2{\beta}(g)}{g}=(\frac{11}{3}N_{c}-\frac{2}{3}N_{f})\frac{\alpha_{s}}{2\pi}= \frac{b{\alpha_{s}}}{{2\pi}}$, where $f_{\phi}$ and $f$ are respectively the dilaton decay constant and a mass scale.

The axion potential (3.8) then becomes:
$$W(a,<\phi>)=\frac{1}{8}\frac{2\beta(g)}{g}(1+\frac{g}{2\beta(g)}\frac{f_{\phi}}{f})<0||G_{\mu\nu}G^{\mu\nu}|0>$$
\begin{equation}
 +\frac{c\left\langle \phi\right\rangle }{4f}N_{c}cos(\frac{1}{N_{C}}(\frac{a}{f_{a}}-\sum\varphi_{i}))
\end{equation}
This expression is similar to VVW model 
and one can get an interesting relation between the quark condensate
and the vaccum expectation value of the dilaton
\begin{equation}
\frac{\left\langle \phi\right\rangle }{f}=\frac{2m_{q}}{N_{c}}N\left\langle 0|\bar{\psi}\psi|0\right\rangle
\end{equation}
Finally, our axion potential can be written as:
$$W(a)=\frac{1}{8}\frac{2\beta(g)}{g}(1+\frac{g}{2\beta(g)}\frac{f_{\phi}}{f})<0||G_{\mu\nu}G^{\mu\nu}|0>$$
\begin{equation}
+m_{q}N_{f}\left\langle 0|\bar{\psi}\psi|0\right\rangle cos(\frac{1}{N_{C}}(\frac{a}{f_{a}}-\sum\varphi_{i}))
\end{equation}
It takes the same form as the potential in$[9]$, where the first term is an
expansion at leading order in $\frac{f\phi}{f}$ ( $\frac{f\phi}{f}\ll1$).
The effective potential given by (3.8) generalizes the VVW potential
which is derived directly in the particular case of the dilaton coupling $F(\phi)=\frac{\phi}{f}$.

\section{Conclusion}
The axion potential in (3.8) depends on the vacuum condensates and
belongs to a class of inflationary potentials describing the QCD vacuum.
It can be derived by invoking the inflaton field {[}11{]} .

In this paper we made an extension of the VVW model by introducing
a massive scalar field. Our idea emerges from the interesting properties
of the dilaton and its crucial role in demestifying some
physical phenomena in particle physics and cosmology.

Through this extended VVW-lagrangian, which takes into account the dilatonic
degrees of freedom, we derived the axion potential like in {[}15
{]}. We showed that the coupling between the dilaton
and the gauge field may be inherited from the nontrivial
structure of Quantum Chromodynamics (QCD) vacuum. In fact, the vacuum
expetation value of the dilaton is propotional to the quark condensate
which appears in the vacuum susceptibility expression whereas
the gluon condensate is proportional to the dilaton mass.

\section*{Acknowledgements}
We would like to thank A. Ihlal and A. Rachidi for their support.

\end{document}